\documentclass[aps,prd,floats,preprint,tightenlines,nofootinbib]{revtex4}
\usepackage{rotating}

\begin{document}
\preprint{ \hbox{hep-ph/0510125} } \vspace*{3cm}

\title{Electroweak Constraints on Effective Theories with $U(2)\times U(1)$ Flavor Symmetry}
\author{Zhenyu Han\footnote{email address:  {\tt zhenyu.han@yale.edu}}}

\affiliation{  \small \sl  Department of Physics, Yale University,
                New Haven, CT  06520\vspace{2.5cm}
            }

\begin{abstract}
In a previous analysis presented in hep-ph/0412166, electroweak
constraints were given on arbitrary linear combinations of a set of
dimension-6 operators. Flavor universality and thus $U(3)^5$ flavor
symmetry were assumed for the operators. In this article, we expand
the analysis to account for the flavor-dependent
theories that distinguish the third generation of fermions from the
light two generations. We still assume flavor universality for the
light two generations to avoid large FCNCs. Consequently, the
$U(3)^5$ flavor symmetry is relaxed to
$[U(2)\times U(1)]^5$, and therefore more operators are added.
We calculate the corrections to electroweak precision observables,
assuming arbitrary coefficients for the operators. The corrections
are combined with the standard model predictions and the
experimental data to obtain the $\chi^2$ distribution as a function
of the operator coefficients. We apply our result to constrain two
flavor-dependent extensions of the standard model: the simplest
little Higgs model and a model with an $SU(2)\times SU(2)\times
U(1)$ gauge group.

\end{abstract}

\maketitle

\newpage

 \section{Introduction}
The standard model (SM) of electroweak physics is generally regarded
as an effective theory with a cut-off far below the Plank scale. The
cut-off is pinned down at the TeV scale if we require that the Higgs
boson mass is not significantly fine-tuned. A variety of extensions
of the SM at the TeV scale have been considered, including
supersymmetry, technicolor, little Higgs and extra dimensions. All
these models predict heavy states that are beyond the current
experimental direct reach (with the Tevatron as a possible
exception). While we are eagerly anticipating the operation of the
LHC to produce and observe these heavy states, they could also leave
their traces in the electroweak precision tests (EWPTs)
\cite{erler+langacker}, by means of quantum corrections. Due to the
remarkable agreement between the SM predictions and the experimental
results, the EWPTs do not tell us which direction of new physics is
particularly promising. Rather, they often put stringent constraints
on the model we are considering.

A model-independent approach to the electroweak constraints is
desirable, in which one needs to calculate the constraints only once
and then is able to apply them to different models. The oblique $S$,
$T$ and $U$ parameter approach \cite{Peskin:1991sw} has exhibited
this merit, and has been effectively used to constrain various
models. However, there exist models with corrections to the SM that
cannot the fully described by the oblique parameters. A general
method that can incorporate both the oblique and non-oblique
corrections is the effective theory approach
\cite{Buchmuller,GW,Han:2004az}.
In this approach, all heavy particles are integrated out and their
effects are manifested in the effective higher order operators
suppressed by powers of the heavy mass scale. For a given order,
there are only a limited number of such operators, to the contrast
of the vast number of possible theoretical extensions. Assuming
arbitrary coefficients for the operators, one calculates the
corrections to the electroweak precision observables (EWPOs) and
compares them with the experimental data. The electroweak
constraints are then obtained in terms of the operator coefficients.

In a previous publication \cite{Han:2004az}, we focused on a set of
dimension-6 operators that are consistent with the SM gauge
symmetry, as well as CP, lepton and baryon number conservation. The
electroweak symmetry breaking was assumed to be linearly realized so
that one or more Higgs doublets were present in the SM particle
spectrum. We also assumed $U(3)^5$ flavor symmetry, in other words,
flavor universality for the operators. This is perhaps the simplest
way to avoid large FCNCs and is well motivated in many theoretical
frameworks. We chose those operators $\mathcal{O}_i$ that could be
tightly constrained and added them to the SM Lagrangian:
\begin{equation}
\mathcal{L}=\mathcal{L}_{SM}+\sum_i a_i\mathcal{O}_i=
\mathcal{L}_{SM}+\sum_i \frac{1}{\Lambda^2_i}\mathcal{O}_i,
\end{equation}
where $\Lambda_i$ has mass dimension one, denoting the energy scale
that suppresses the operator $\mathcal{O}_i$. We calculated the
corrections to the EWPOs linearly in $a_i$ and obtained the bounds
on arbitrary linear combinations of these operators. The results
have been used to constrain the mass
of a generic $Z'$ gauge boson \cite{Han:2004az}, and the parameter
spaces of a variety of little Higgs models \cite{Han:2005dz}.

Nevertheless, small FCNCs can be achieved without assuming flavor
universality. Technically, each of the $U(3)$ symmetries can be
relaxed to $U(1)^3$ with one $U(1)$ corresponding to one generation,
as long as each generation is a mass eigenstate. With the $U(1)^3$
symmetry, operators involving different generations can have
different coefficients, and still do not contribute to FCNCs.
However, from the model-building point of view, it is unnatural to
assume that the effective operators only contain mass eigenstates.
In fact, in many models, operators are obtained by
integrating out heavy gauge bosons, and involve gauge eigenstates
instead. But natural flavor-dependent models at TeV scale are still
possible, as long as universality is conserved for the first two
generations and only the third generation is treated differently. 
This is mainly due to two reasons. First, the mass and gauge 
eigenstates of the third generation of quarks are quasi-aligned,
namely, they are almost identical up to small rotations of order 
$\mathcal{O}(\lambda^2)$, where 
$\lambda\simeq 0.22$ is the Wolfenstein parameter in the CKM matrix.
This is to the contrast of the first two generations of quarks, where the 
mixing is of order $\mathcal{O}(\lambda)$. Second, the most stringent 
experimental constraints on flavor violation come from the Kaon system, 
which can be avoided by assuming universality for the first two 
generations. The flavor changing experiments involving the 
third generation currently have limited statistics, and do 
not rule out TeV scale flavor physics \cite{Agashe:2005hk}.

Given the above observation,  we extend the analysis in
Ref.~\cite{Han:2004az} on electroweak constraints to
incorporate flavor-dependent operators. We relax each of the $U(3)$
symmetries to $U(2)\times U(1)$, with the $U(2)$ reflecting the
universality of the first two generations. The $U(1)$ corresponds to
flavor conservation in the third generation. We do not attempt to
analyze the bounds from flavor changing measurements, which is a
rich and interesting topic in itself. See Ref.~\cite{Agashe:2005hk}
for a recent analysis. Rather, we will use the flavor-conserving
EWPTs to constrain arbitrary linear combinations of all relevant
flavor-dependent dimension-6
operators. We will use mass eigenstates as the basis of the
operators. However, due to the universality in the first two
generations and the quasi-alignment in the third generation
discussed in the previous paragraph, the results apply as well to gauge
eigenstates, and in general, to any basis that differs by only small
rotations. Of course, since the symmetry is relaxed, there are more
operators in this case than in the flavor-independent case. We
enumerate and analyze the constraints for these operators in the
next section. In Sec.~\ref{sec:app}, the results are used to put
bounds on two flavor-dependent models. Sec.~\ref{sec:summary}
contains the summary and discussion.

 \section{Operators and constraints}
 \label{sec:ops}

Assuming $U(3)^5$ flavor symmetry, we obtained 21 dimension-6
operators that are relevant to EWPTs in Ref.~\cite{Han:2004az}.
Using $W^a_{\mu \nu}$ and $B_{\mu \nu}$ to denote the $SU(2)_L$ and
$U(1)_Y$ gauge boson field strength, $l,q$ the left handed leptons and
quarks, $e,u,d$ the right handed leptons and quarks, and $h$ the
Higgs doublet, we list them below.
\begin{enumerate}
\item{Operators modifying gauge boson propagators:}
\begin{eqnarray}
 &&
 O_{W\!B}=(h^\dagger \sigma^a h) W^a_{\mu \nu} B^{\mu \nu}, \  \  \   O_h = | h^\dagger D_\mu h|^2;
 \label{op:ST}
\end{eqnarray}
\item{Four-fermion operators:}
\begin{eqnarray}
 &&
 O_{ll}^s=\frac{1}{2} (\overline{l} \gamma^\mu l) (\overline{l} \gamma_\mu l), \ \ \
 O_{ll}^t=\frac{1}{2} (\overline{l} \gamma^\mu \sigma^a l) (\overline{l} \gamma_\mu \sigma^a l),
 \nonumber\\&&
 O_{lq}^s= (\overline{l} \gamma^\mu l) (\overline{q} \gamma_\mu q), \ \ \
 O_{lq}^t= (\overline{l} \gamma^\mu \sigma^a l) (\overline{q} \gamma_\mu \sigma^a q),
 \nonumber\\ &&
 O_{le}= (\overline{l} \gamma^\mu l) (\overline{e} \gamma_\mu e),  \ \ \
 O_{qe}=(\overline{q} \gamma^\mu q) (\overline{e} \gamma_\mu e),
 \nonumber\\ &&
 O_{lu}= (\overline{l} \gamma^\mu l) (\overline{u} \gamma_\mu u),  \ \ \
 O_{ld}= (\overline{l} \gamma^\mu l) (\overline{d} \gamma_\mu d),
 \nonumber\\ &&
 O_{ee}=\frac{1}{2} (\overline{e} \gamma^\mu e) (\overline{e} \gamma_\mu e), \ \ \
 O_{eu}=(\overline{e} \gamma^\mu e) (\overline{u} \gamma_\mu u),  \ \ \
 O_{ed}=(\overline{e} \gamma^\mu e) (\overline{d} \gamma_\mu d);\label{op:4f}
\end{eqnarray}
\item{Operators modifying gauge-fermion couplings:}
\begin{eqnarray}
 &&
 O_{hl}^s = i (h^\dagger D^\mu h)(\overline{l} \gamma_\mu l) + {\rm h.c.}, \ \ \
 O_{hl}^t = i (h^\dagger \sigma^a D^\mu h)(\overline{l} \gamma_\mu \sigma^a l)+ {\rm h.c.},
 \nonumber\\ &&
 O_{hq}^s = i (h^\dagger D^\mu h)(\overline{q} \gamma_\mu q)+ {\rm h.c.}, \ \ \
 O_{hq}^t = i (h^\dagger \sigma^a D^\mu h)(\overline{q} \gamma_\mu \sigma^a q)+ {\rm h.c.},
 \nonumber\\ &&
 O_{hu} = i (h^\dagger D^\mu h)(\overline{u} \gamma_\mu u)+ {\rm h.c.}, \ \ \
 O_{hd} = i (h^\dagger D^\mu h)(\overline{d} \gamma_\mu d)+ {\rm h.c.},
 \nonumber\\ &&
 O_{he} = i (h^\dagger D^\mu h)(\overline{e} \gamma_\mu e)+ {\rm h.c.}\,;\label{op:hf}
\end{eqnarray}
\item{Operator modifying the triple-gauge couplings:}
\begin{equation}
O_W=\epsilon^{abc} \, W^{a \nu}_{\mu} W^{b\lambda}_{\nu} W^{c \mu}_{\lambda}.\label{op:W}
\end{equation}
\end{enumerate}
Operators containing fermions are understood to be summed over
flavor indices, consistent with the $U(3)^5$ symmetry. Note that the
operators $O_{WB}$ and $O_h$ in Eq.~(\ref{op:ST}) correspond to the
oblique $S$ and $T$ parameters. $O_{WB}$ also contains a term that
modifies the triple gauge boson couplings. Four-fermion operators containing
only quarks are not inucluded in the list because they cannot be
tightly constrained by available data.

With the relaxed $[U(2)\times U(1)]^5$ symmetry, the operators have
similar structures. We only need to single out the third generation
from the light two. By slightly abusing the notation, from now
on, we will use $l,q,e,u$ and $d$ to represent only the first two
generations of fermions. By this change of definition, the 
operators in Eqs.~(\ref{op:ST})$-$(\ref{op:W}) are still present in
the relaxed-symmetry case. Three of them, $O_{W\!B}$, $O_h$ and
$O_W$ do not involve fermions. They are the same as in the $U(3)^5$
case. The other operators are now understood to be summed over only
the first two generations, reflecting the $U(2)^5$ symmetry.

We then turn to the operators involving the third generation. We use
$L,Q,\tau,t,b$ to denote the third generation counterparts of
$l,q,e,u,d$, respectively. The relavant operators are listed below in
Eqs.~(\ref{op:4f2}) and (\ref{op:hf2}). We only consider those
operators that can be tightly constrained by the current
experimental data. Therefore, four-fermion operators involving
only the third generation have been omitted. We have also omitted
the operator $O_{ht}$ that modifies the coupling between the $Z$ boson
and the top quark, as well as the operators $O_{lt}$ and $O_{et}$,
since the top quark is not involved in any of the
EWPTs.
\begin{enumerate}
\item{Four-fermion operators:}
\begin{eqnarray}
  &&
  O_{lL}^s= (\overline{l} \gamma^\mu l) (\overline{L} \gamma_\mu L), \ \ \
  O_{lL}^t= (\overline{l} \gamma^\mu \sigma^a l) (\overline{L} \gamma_\mu \sigma^a L),
  \nonumber\\ &&
  O_{lQ}^s= (\overline{l} \gamma^\mu l) (\overline{Q} \gamma_\mu Q), \ \ \
  O_{lQ}^t= (\overline{l} \gamma^\mu \sigma^a l) (\overline{Q} \gamma_\mu \sigma^a Q),
  \nonumber\\ &&
  O_{Le}= (\overline{L} \gamma^\mu L) (\overline{e} \gamma_\mu e),  \ \ \
  O_{l\tau}= (\overline{l} \gamma^\mu l) (\overline{\tau} \gamma_\mu \tau),
  \nonumber\\ &&
  O_{Qe}=(\overline{Q} \gamma^\mu Q) (\overline{e} \gamma_\mu e),\ \ \
  O_{lb}= (\overline{l} \gamma^\mu l) (\overline{b} \gamma_\mu b),
  \nonumber\\ &&
  O_{e\tau}=(\overline{e} \gamma^\mu e) (\overline{\tau} \gamma_\mu \tau), \ \ \
  O_{eb}=(\overline{e} \gamma^\mu e) (\overline{b} \gamma_\mu b);
  \label{op:4f2}
\end{eqnarray}
\item{Operators modifying gauge-fermion couplings:}
\begin{eqnarray}
  &&
  O_{hL}^s = i (h^\dagger D^\mu h)(\overline{L} \gamma_\mu L) + {\rm h.c.}, \ \ \
  O_{hL}^t = i (h^\dagger \sigma^a D^\mu h)(\overline{L} \gamma_\mu \sigma^a L)+ {\rm h.c.},
  \nonumber\\ &&
  O_{hQ}^s = i (h^\dagger D^\mu h)(\overline{Q} \gamma_\mu Q)+ {\rm h.c.}, \ \ \
  O_{hQ}^t = i (h^\dagger \sigma^a D^\mu h)(\overline{Q} \gamma_\mu \sigma^a Q)+ {\rm h.c.},
  \nonumber\\ &&
  O_{h\tau} = i (h^\dagger D^\mu h)(\overline{\tau} \gamma_\mu \tau)+ {\rm h.c.},\ \ \
  O_{hb} = i (h^\dagger D^\mu h)(\overline{b} \gamma_\mu b)+ {\rm h.c.}.
  \label{op:hf2}
\end{eqnarray}
\end{enumerate}

The complete list of 37 operators relevant to our analysis are given
by Eqs.~(\ref{op:ST})$-$(\ref{op:hf2}). Assuming arbitrary
coefficients $a_i$ for operators $O_i$, we have calculated to the
linear order in $a_i$ the corrections to EWPOs from these operators.
The validity of the linear approximation will be discussed later in
this section. The corrections are compared with the experimental
data and the total $\chi^2$ is obtained as a function of $a_i$.
The experimental data contain all precisely measured observables relevant
to our analysis, including the $W$ boson mass, observables from atomic
parity violation, deep inelastic scattering and $Z$-pole experiments,
and fermion and $W$ boson pair production data from LEP 2. Since
the corrections are assumed to be linear in $a_i$, the $\chi^2$ is
quadratic:
\begin{equation}
\chi^2(a_i)=\chi^2_{SM}+ a_i \hat{v_i} + a_i {\mathcal M}_{ij} a_j. \label{chi2}
\end{equation}
In the above equation, $\chi^2_{SM}$ is the value of $\chi^2$ when
all $a_i$ are set to zero. The vector $\hat{v_i}$ and the matrix
$\mathcal M$ are our main results. Their numerical values are given
in Ref.~\cite{code}. The calculations of the corrections and the
$\chi^2$ distribution are very similar to the previous work in
Ref.~\cite{Han:2004az}. Thus we refer readers to
Ref.~\cite{Han:2004az} and references therein, and the Mathematica
code \cite{code} for the details. Here we only comment on the
results.

Unlike the $U(3)^5$ case, not all of the operators can
be independently constrained by the current data. There exist ``flat directions'' in which
some combinations of the operators have vanishing corrections to
all of the EWPOs and thus can not be constrained. This is due to the
increase in the number of operators and the lack of relevant
measurements. To remove as many as possible flat directions, we have
added a few observables that were omitted in the previous analysis.
These observables were omitted because they are less precisely
measured than others that could constrain the same operators, due to
the large $U(3)^5$ symmetry.

The added observables are mainly the total cross-sections and
asymmetries in the processes $e^+e^-\rightarrow b\overline b$ and
$e^+e^-\rightarrow c\overline c$ at LEP 2 \cite{lep2}. They are
useful for removing flat-directions for 4-fermion operators
containing 2 leptons and 2 quarks. Without these data, there are 5
flat directions involving such operators. This is what one would
expect: because the third generation is separated from the first
two, there are five new 4-fermion operators containing heavy quarks:
$O_{lQ}^s,O_{lQ}^t,O_{Qe},O_{lb}$ and $O_{eb}$. However the only
data in the previous analysis that can constrain these operators is
the total hadronic cross-section for $e^+e^-$ scattering measured at
LEP 2, and the constraints from these data are still correlated with
the constraints on operators containing the light quarks. After
adding the $e^+e^-\rightarrow b\overline b$ and $e^+e^-\rightarrow
c\overline c$ data, 4 out of the 5 flat directions are removed. The
remaining flat direction comes from the fact that the operators $O_{lQ}^s$ and
$O_{lQ}^t$ contribute the same to the $e^+e^-\rightarrow b\overline
b$ process:
\begin{equation}
a_{lQ}^s=-a_{lQ}^t, \quad\mbox{all other } a_i=0.\label{flat}
\end{equation}

Less important observables added to the analysis are the asymmetries
for the strange quark: $A_s$ and $A_{FB}^{0,s}$
\cite{erler+langacker} measured at the Z-pole. They are much less
precisely measured than the asymmetries for the bottom quark: $A_b$
and $A_{FB}^{0,b}$, which would constrain the same operators if
$U(3)^5$ were assumed. It turns out that adding these observables
does not remove any of the flat-directions, but helps to refine the
constraints.

Besides the one in Eq.~(\ref{flat1}), there are three other flat
directions:
\begin{eqnarray}
a_{Le}=-a_{l\tau}, \quad\mbox{all other } a_i=0;\nonumber\\
a_{lL}^s=-a_{lL}^t,\quad\mbox{all other } a_i=0;\nonumber\\
a_{hQ}^s=-a_{hQ}^t, \quad\mbox{all other } a_i=0.\label{flat1}
\end{eqnarray}
These flat directions reflect the lack of EWPOs involving top pair
production and $\nu_\tau$-nucleon, $\nu_\tau$-lepton scattering.

Besides the exact flat directions, there also exist
``weakly-bounded'' directions, which means that these directions
cannot be tightly constrained by the available data. To illustrate
this point, we consider the eigenvalues and the corresponding
eigenvectors of the matrix $\mathcal{M}$ in Eq.~(\ref{chi2}). By
doing so, the correlations between different directions are removed.
Each eigenvector represents a direction that can be independently
constrained. The corresponding eigenvalue is the $1 \sigma$
constraint on the scale $\Lambda_i$ for this direction. Numerically,
four of the eigenvalues are zeroes, corresponding to the four flat
directions discussed above. The next few smallest eigenvalues are
190, 210, 270, 280 and 580 GeV. Bounds on $\Lambda_i$ this small
cannot be taken literally. This is because when calculating the
corrections to EWPOs, we only work to the linear order in the
coefficients $a_i$, corresponding to the linear order in
$v^2/\Lambda_i^2$ or $E^2/\Lambda_i^2$. Here $v=246$ GeV, is the
electroweak symmetry breaking scale, and $E$ is the energy scale of
a given measurement, which goes up to about 200 GeV for the LEP 2
measurements. Since the smallest nonzero eigenvalues mentioned above
are close to $v$ or $E$, the second order corrections in $a_i$ are
of the similar size and thus cannot be neglected. To obtain precise
bounds one has to include these second order corrections, as well as
corrections from higher order operators beyond dimension-6. In this
case, a model-by-model approach would be more convenient.

The existence of the flat and weakly-bounded directions is not a
deficiency of our method. Rather, it reflects the fact that not
enough measurements are available to tightly constrain all
independent operators. In principle, these directions could also be
constrained if new or more precise measurements were available. In
practice, the chance is rare for the coefficients of the effective
operators in a model to coincide with the flat or weakly-bounded
directions. Moreover, one can easily check if the obtained bounds
are large enough to ignore the higher order corrections. Usually a
bound on $\Lambda_i$ about 1 TeV is large enough, because higher
order corrections are suppressed by extra powers of
$v^2/\Lambda_i^2$ or $E^2/\Lambda_i^2$, which are negligible for
$\Lambda_i\gtrsim 1$ TeV. Even if we obtain bounds that are much
smaller than 1 TeV and thus not precise, we will still be certain
that the exact bounds will not be much tighter. Often, learning that
a model is not tightly constrained by EWPTs is interesting enough.

In the following section, we apply our general result to obtain the
bounds for two flavor-dependent models. We will first integrate out
the heavy degrees of freedom and obtain the operator coefficients
$a_i$ as functions of the parameters in the theory. Then it is
straightforward to obtain the constraints by substituting $a_i$ in
the $\chi^2$ distribution, Eq.~(\ref{chi2}).

 \section{Applications}
 \label{sec:app}
\subsection{The Simplest Little Higgs Model \cite{simplest}}
The first model we consider is the simplest little Higgs model
discussed in Ref.~\cite{simplest}. Little Higgs models
\cite{LHmodels,LHreview,simplest} are a set of models that aim at
solving the fine-tuning problem associated with the Higgs mass, by
introducing extra gauge and matter fields with masses at TeV scale.
The simplest little Higgs model extends the electroweak gauge group
to $SU(3)\times U(1)$. To each fermion doublet, a heavy fermion is
added to form a triplet of the $SU(3)$ group. While the gauge group
is fixed, there is still freedom to choose from different charge
assignments for the fermions. In Ref.~\cite{simplest}, two such
charge assignments are discussed, which are called model 1 and model
2. In model 1, the three generations of fermions are assigned the
same quantum numbers. The only flavor-dependent effect comes from
the Yukawa couplings. Thus we obtain (approximately)
flavor-universal operators and the constraints have been given in
Ref.~\cite{Han:2005dz}. We focus on model 2 in this article.  In
model 2, the third generation of quark triplet is assigned different
quantum numbers from the light two generations. The charge
assignments for the three generations of lepton triplets, as well as
lepton and quark singlets
are the same. Thus we expect a $U(3)^4\times U(2)\times U(1)$ group
as the flavor symmetry. This symmetry is larger than $[U(2)\times
U(1)]^5$ we assumed for the analysis, and is manifested by the
relations between the operator coefficients, as described below. The
notation follows Ref.~\cite{simplest}.

We split the effective operators to two parts. The first part comes
from integrating out the heavy gauge bosons and the coefficients
$a_i$ can be expressed in a condensed form
\begin{eqnarray}
a_h&=&-\frac{9}{4F^2}\frac{(1-\frac23x^2)^2}{(3+x^2)^2},\nonumber\\
a_{hf}^s&=&\frac{9}{4F^2}\frac{1-\frac23x^2}{(3+x^2)^2}(\sqrt{3}T^{8f}+x^2
Y_x^f),\nonumber\\
a_{ff'}^s&=&-\frac{9}{2F^2}\frac{1}{(3+x^2)^2}(\sqrt{3}T^{8f}+x^2
Y^f_x)(\sqrt{3}T^{8f'}+x^2Y^{f'}_x),\label{su3op1}
\end{eqnarray}
where $x=g_x/g$, $g_x^2={3g^2g'^2}/(3g^2-g'^2)$, and $F^2=f_1^2+f_2^2$ 
depicting the heavy mass scale. The quantum
numbers for the fermions $f=q,l,u,d,e,Q,L,b,\tau$ are given by
\begin{eqnarray}
T^{8f}&=&-\frac{1}{2\sqrt{3}},\ \frac{1}{2\sqrt{3}},\ 0,\ 0,\ 0,\ \frac{1}{2\sqrt{3}},\ \frac{1}{2\sqrt{3}},\ 0,\ 0;\\
Y_x^f&=&0,\ -\frac13,\ \frac23,\ -\frac13,\ -1,\ \frac13,\
-\frac13,\ -\frac13,\ -1.
\end{eqnarray}

The second part of the operators comes from integrating out the heavy
fermions. The Yukawa couplings come from the following Lagrangian.
\begin{equation}
\lambda_1^uu^{c3}_1\Phi^\dag_1\Psi_{Q^3}+\lambda_2^uu^{c3}_2\Phi^\dag_2\Psi_{Q^3}+
\lambda_1^dd^{c1,2}_1\Phi_1\Psi_{Q^{1,2}}+\lambda_2^dd^{c1,2}_2\Phi_2\Psi_{Q^{1,2}}+
\lambda^nn^c\Phi_1^\dag\Psi_L+\ldots
\end{equation}
We have omitted the terms that mix the third generation with the
first two generations, such as $u^{c1,2}\Phi_1^\dag\Psi_{Q^3}$,
since the mixing is very small. In this case, the couplings
$\lambda^d_{1,2}$ are $2\times2$ matrices. In order to avoid large
FCNCs, similar to Ref.~\cite{KS}, we take one of them to be
proportional to the identity matrix and of order one, and the other
one approximately the down-type Yukawa matrix for the first two
generations and thus much smaller in magnitude. When integrating out
the heavy fermions, we can then neglect the smaller $\lambda^d$. In
this way, we obtain operators that conserve the $U(2)$ symmetry for
the light two generations:
 \begin{eqnarray}
  a_{hl}^s&=&-a_{hl}^t=\frac14\frac{f_2^2}{F^2 f_1^2},\nonumber\\
  a_{hq}^s&=&a_{hq}^t=\left\{
     \begin{array}{c}
      \frac14\frac{f_1^2}{F^2 f_2^2}\quad(\lambda_1^d\ll\lambda_2^d)\\
      \frac14\frac{f_2^2}{F^2 f_1^2}\quad(\lambda_2^d\ll\lambda_1^d)
     \end{array}\right..\label{su3op2}
 \end{eqnarray}
The equality
between $a_{hq}^s$ and $a_{hq}^t$ is a consequence of the fact that
the heavy quarks mix with only the down-type quarks in the light two
generations. For the third generation, the heavy quark mixes with
the top quark and we get $a_{hQ}^s=-a_{hQ}^t$ similarly. According
to Eq.~(\ref{flat1}), this is a flat direction, so we simply set them
to zero.

\begin{figure}
\begin{center}
   \includegraphics[width=\textwidth]{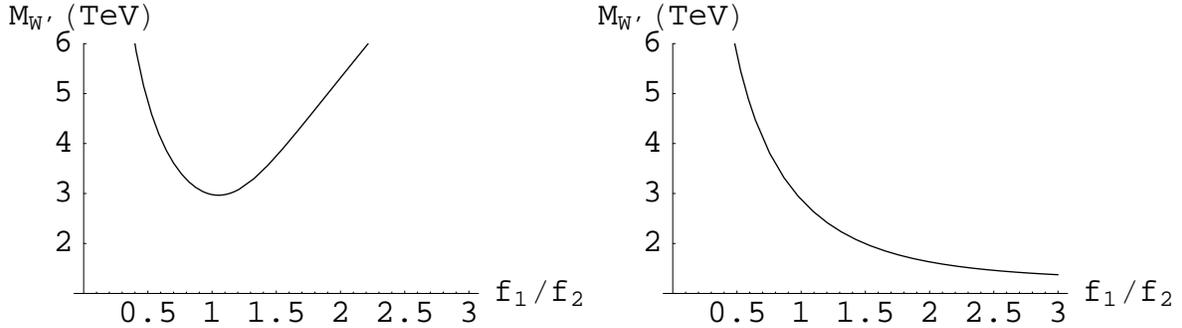}
\end{center}
  \caption{Lower bounds at 95\% CL on $M_{W'}$ as a function of $f_1/f_2$ in the simplest
little Higgs model. Left:
 $\lambda_1^d\ll\lambda_2^d$; right: $\lambda_2^d\ll\lambda_1^d$.}
\label{fig:su3}
\end{figure}

Combining Eqs.~(\ref{su3op1}) and (\ref{su3op2}) and substituting
them to the $\chi^2$ distribution, we can obtain the bounds on the
scale $F$. For comparison with the bounds on model 1, given in
Ref.~\cite{Han:2005dz}, we translate the bounds on $F$ to the bounds
on the mass of the $W'$ gauge boson by the relation
\begin{equation}
M^2_{W'}=g^2F^2/2.
\end{equation}
The 95\% confidence level (CL) bounds on $M_{W'}$ as a function of
$f_1/f_2$ are shown in Figure \ref{fig:su3}. One of the main
motivations to consider the constraints is to estimate the
associated fine-tuning. The heavy gauge bosons are introduced to
cancel the quadratically divergent corrections to the Higgs boson
mass-squared from the SM gauge boson loops. In order to avoid more
than 10\% fine-tuning, the $W'$ boson mass should be smaller than
about 5 TeV \cite{LHreview}. We see from Figure \ref{fig:su3} that
$M_{W'}<5$ TeV is allowed for a large portion of the parameter
space. It is interesting that for the $\lambda_2^d\ll\lambda_1^d$
case, the bounds even go down to less than 1.5 TeV on the $f_1>f_2$
side.

\subsection{An $SU(2)\times SU(2)\times U(1)$ model \cite{Morrissey:2005uz}}
The second model is an $SU(2)\times SU(2)\times U(1)$ model
discussed in Ref.~\cite{Morrissey:2005uz}. The electroweak gauge
group is enlarged to $SU(2)_1\times SU(2)_2\times U(1)_Y$. The
$U(1)_Y$ coincides with the SM $U(1)_Y$ group. The SM $SU(2)_L$
group is the diagonal subgroup of $SU(2)_1\times SU(2)_2$. The
authors of Ref.~\cite{Morrissey:2005uz} mainly focus on the instanton
effects associated with the larger gauge group, but electroweak
constraints on the model are also given. Only the data of $Z$-pole
measurements and the $W$ boson mass are used in their analysis. In
this subsection, we show that it is straightforward to obtain
constraints from a much larger set of observables using our
approach.

As in the previous model, we first obtain the effective operators.
The third generation of fermion doublets are assigned different
quantum numbers from the light two generations. In our notation, the
quantum numbers under $SU(2)_1\times SU(2)_2\times U(1)_Y$ are
\begin{equation}
Q:(2,1)_{1/6},\quad  L:(2,1)_{-1/2},\quad q: (1,2)_{1/6},\quad l:(1,2)_{-1/2}.
\end{equation}
The $SU(2)_L$ singlet fields are singlets under both $SU(2)_1$ and $SU(2)_2$,
with the $U(1)_Y$ charges the same as the SM hypercharges.

We have two choices for the Higgs doublet quantum numbers. In
Ref.~\cite{Morrissey:2005uz}, they are called the ``heavy'' case
$h=(2,1)_{1/2}$ and the ``light'' case $h=(1,2)_{1/2}$.

The $SU(2)_1\times SU(2)_2$ gauge group is broken to $SU(2)_L$ by
the vacuum expectation value of a bidoublet scalar
$\langle\Sigma\rangle=\mbox{diag}\{u,u\}.$ The gauge coupling $g$ of
the unbroken $SU(2)$ is related to the gauge couplings of $SU(2)_1$
and $SU(2)_2$ as
\begin{equation}
g=\frac{g_1g_2}{\sqrt{g_1^2+g_2^2}}.
\end{equation}
For convenience, we define
\begin{equation}
c=\frac{g_1}{\sqrt{g_1^2+g_2^2}},\quad s=\frac{g_2}{\sqrt{g_1^2+g_2^2}}.
\end{equation}

The symmetry breaking yields three heavy gauge bosons $Z',W'^\pm$
with masses
\begin{equation}
 M^2_{Z'}=M^2_{W'^\pm}=(g_1^2+g_2^2)u^2.\label{su2mass}
\end{equation}
We integrate out these gauge bosons in both the light and heavy cases and obtain the following operator coefficients

Light case:
\begin{eqnarray}
&&a^t_{ll}=a^t_{lq}=a^t_{hl}=a^t_{hq}=-\frac{1}{4u^2}s^4,\nonumber\\
&&a^t_{lL}=a^t_{lQ}=a^t_{hL}=a^t_{hQ}=\frac{1}{4u^2}s^2c^2.\label{lightops}
\end{eqnarray}

Heavy case:
\begin{eqnarray}
&&a^t_{ll}=a^t_{lq}=-\frac{1}{4u^2}s^4,\nonumber\\
&&a^t_{lL}=a^t_{lQ}=a^t_{hl}=a^t_{hq}=\frac{1}{4u^2}s^2c^2,\nonumber\\
&&a^t_{hL}=a^t_{hQ}=-\frac{1}{4u^2}c^4.\label{heavyops}
\end{eqnarray}

\begin{figure}
\begin{center}
   \includegraphics[width=0.6\textwidth]{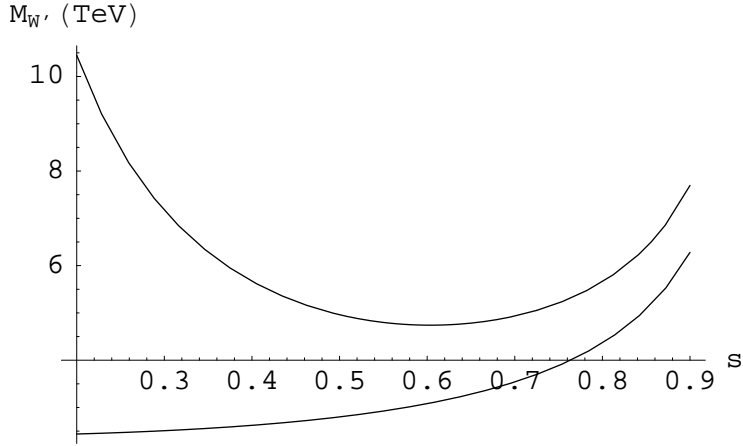}
\end{center}
  \caption{Lower bounds at 95\% CL on $M_{W'}$ as a function of $s$ in the $SU(2)\times SU(2)\times U(1)$ model.
The upper curve corresponds to the heavy case and the lower curve corresponds to the light case.}
\label{fig:su2}
\end{figure}

The authors of Ref.~\cite{Morrissey:2005uz} have calculated the
corrections to the Z-pole observables and the $W$ boson mass in
terms of the parameters $u$ and $s$. Given the simple form of
operator coefficients in Eqs.~(\ref{lightops}) and (\ref{heavyops}),
it is straightforward to reproduce their results from our general
calculations with arbitrary operator coefficients. In fact, it does
not take more effort to substitute the coefficients in
Eq.~(\ref{chi2}) to obtain more comprehensive constraints. We
translate the bounds on $u$ to the bounds on the physical mass
$M_{W'}$ ($M_{Z'}$) using the relation (\ref{su2mass}). The 95\% CL
bounds using all data in our analysis are shown 
in Figure \ref{fig:su2}, corresponding 
to $\Delta\chi^2=3.84$. If we restrict the data to those 
used in Ref.~\cite{Morrissey:2005uz}, we obtain curves of 
similar shapes for the bounds. But the bounds are generally lower,
with the largest difference about 1 TeV for the heavy case and 
2 TeV for the light case. Note that in Figure \ref{fig:su2}, the bounds
for the heavy case are quite stringent and always tighter than 
those for the light case. But in both cases, 
the heavy gauge bosons will have chances to be observed at the LHC. 

\section{Summary and discussion}
\label{sec:summary}
We have analyzed constraints from EWPTs on flavor-dependent
extensions of the SM at TeV scale. Our results are model-independent
in the sense that the constraints are given in terms of the bounds
on arbitrary linear combinations of dimension-6 effective
operators. The analysis is an extension to
the previous work presented in Ref.~\cite{Han:2005dz}. In
Ref.~\cite{Han:2005dz}, $U(3)^5$ flavor symmetry is assumed for
the matter fields, leaving no room for flavor-dependent effects. In
this article, we have relaxed the flavor symmetry to $[U(2)\times
U(1)]^5$, with the $U(1)$'s corresponding to the third generation.
Constraints from FCNCs on the first two generations are much more
stringent than the third generation, indicating that
flavor-dependent physics for the first two generations have to arise
at a much higher scale than 1 TeV. Therefore we still assume $U(2)$
symmetry for the first two generations.

There are 16 more operators that are relevant to EWPTs than in the
previous analysis. There exists a potential problem with so many
operators, that is, some operator combinations (which we call flat
directions) can not be constrained by current experimental data 
because their net corrections to
all EWPOs vanish. This is true even after we have added a few
measurements that were omitted previously due to their low
precision. However, in practice, the flat directions are almost
never a problem, since there are usually much fewer parameters than
effective operators in a given model. On the other hand, one might
utilize the flat directions to avoid strong electroweak constraints
when building a model. Since the flat directions given in
Eqs.~(\ref{flat}) and (\ref{flat1}) all involve the third
generation, it will be easier to avoid the constraints if the new
physics couples exclusively to the third generation.

We have calculated the $\chi^2$ distribution in terms of the
operator coefficients. To constrain a given model, the only thing
one still needs to do is integrating out the heavy degrees of
freedom to obtain the operator coefficients in term of the model
parameters. We have applied this procedure to two flavor-dependent
models. The first one is the simplest little Higgs, model 2
\cite{simplest}. We found that the bounds on the heavy $W'$ gauge
boson mass are even lower than that in model 1, indicating this
model is a good solution to the Higgs mass fine-tuning problem. The
second model we have considered is a model with the SM gauge group
enlarged to $SU(2)\times SU(2)\times U(1)$, discussed in
Ref.~\cite{Morrissey:2005uz}. The authors of
Ref.~\cite{Morrissey:2005uz} obtained the electroweak constraints
for this model from a subset of all available EWPOs. We have shown
that it is straightforward to obtain more comprehensive constraints
utilizing our result. Generally speaking, our result can be used to 
efficiently constrain any TeV scale flavor-dependent physics with $U(2)\times U(1)$
flavor symmetry.  

\section*{Acknowledgments}
The author would like to thank Witold Skiba for numerous helpful
discussions, and Yuval Grossman for comments and suggestions on the manuscript.
 This research was supported in part by the US
Department of Energy under grant  DE-FG02-92ER-40704.

\end{document}